\documentclass[sigconf]{acmart}

\usepackage{booktabs}
\usepackage{tabularx}
\usepackage{makecell}
\usepackage{listings}
\usepackage{newunicodechar}
\newcolumntype{P}[1]{>{\raggedright\arraybackslash}p{#1}}
\newcolumntype{L}{>{\raggedright\arraybackslash}X}

\newunicodechar{β}{\ensuremath{\beta}}
\newunicodechar{α}{\ensuremath{\alpha}}
\newunicodechar{Δ}{\ensuremath{\Delta}}
\newunicodechar{χ}{\ensuremath{\chi}}
\newunicodechar{−}{\ensuremath{-}}
\newunicodechar{≤}{\ensuremath{\leq}}
\newunicodechar{×}{\ensuremath{\times}}
\newunicodechar{²}{\textsuperscript{2}}
\newunicodechar{₀}{\ensuremath{_0}}
\newunicodechar{₁}{\ensuremath{_1}}

\setcopyright{none}
\renewcommand\footnotetextcopyrightpermission[1]{}
\acmConference{}{}{}
\acmDOI{}
\acmISBN{}

\begin{document}

\title{Modeling of Therapist–Client Dynamics in Psychotherapy Using LLM-Based Assessments: An Empirical Study}

\author{Angela Chen}
\affiliation{%
  \institution{Carnegie Mellon University}
  \city{Pittsburgh}
  \state{PA}
  \country{USA}}

\author{Siwei Jin}
\authornote{Both authors contributed equally to this research.}
\affiliation{%
  \institution{Carnegie Mellon University}
  \city{Pittsburgh}
  \state{PA}
  \country{USA}}

\author{Canwen Wang}
\authornotemark[1]
\affiliation{%
  \institution{Carnegie Mellon University}
  \city{Pittsburgh}
  \state{PA}
  \country{USA}}

\author{Holly Swartz}
\affiliation{%
  \institution{University of Pittsburgh}
  \city{Pittsburgh}
  \state{PA}
  \country{USA}}

\author{Tongshuang Wu}
\affiliation{%
  \institution{Carnegie Mellon University}
  \city{Pittsburgh}
  \state{PA}
  \country{USA}}

\author{Robert E Kraut}
\affiliation{%
  \institution{Carnegie Mellon University}
  \city{Pittsburgh}
  \state{PA}
  \country{USA}}

\author{Haiyi Zhu}
\authornote{Corresponding author.}
\email{haiyiz@andrew.cmu.edu}
\affiliation{%
  \institution{Carnegie Mellon University}
  \city{Pittsburgh}
  \state{PA}
  \country{USA}}

\renewcommand{\shortauthors}{Chen et al.}

\begin{abstract}
\textbf{Background:} Psychotherapy is a primary treatment for many mental health conditions, yet the dynamic interplay among therapist behaviors, client responses, and the therapeutic relationship remains difficult to disentangle at scale. Traditional process research relies on labor-intensive human coding, limiting the ability to model moment-to-moment therapeutic dynamics across large datasets.

\textbf{Objective:} This study aimed to develop and validate a computational framework for modeling therapist--client interaction processes and to examine how therapist behaviors and relational qualities relate to client self-disclosure and emotional expression in psychotherapy sessions.

\textbf{Methods:} We developed automated assessment pipelines using large language models (LLMs) to measure therapist behaviors (empathy, exploration), relational quality (rapport), and client outcomes (self-disclosure, self-directed negative emotion, outward-directed negative emotion). Model-generated scores were validated against human annotations. We then applied these measures to approximately 2,000 hours of psychotherapy transcripts from the Alexander Street corpus. This yielded an estimation sample of 119,233 client utterances from 1,573 sessions and 37 clients. Because utterances are nested within sessions and sessions within clients, we estimated a saturated multivariate structural equation model (SEM) with standard errors clustered at the session level. Results were confirmed in a multilevel regression model with random intercepts for session and client.

\textbf{Results:} Following comparative validation of six candidate language models, automated measures showed good to excellent agreement with human annotations (ICC = 0.60--0.81), with rapport (ICC = 0.81), self-disclosure (ICC = 0.78), empathy components (ICC = 0.60--0.75), and emotion measurements (ICC = 0.66--0.80). SEM analyses revealed that therapist empathy and exploration were each associated with increased client disclosure (empathy β=0.086, 95\% CI 0.078 to 0.095; exploration β=0.086, 95\% CI 0.078 to 0.094; both P\textless.001) and increased expression of self-directed negative emotions, comprising sadness, anxiety, fear, and depression (empathy β=0.118, 95\% CI 0.110 to 0.127; exploration β=0.078, 95\% CI 0.070 to 0.087; both P\textless.001). Both behaviors were also associated with reduced expressions of enjoyment (empathy β=−0.037; exploration β=−0.041; both P\textless.001). Surprisingly, rapport in the prior session showed no association with client self-disclosure or negative emotion expression.

\textbf{Conclusions:} This study demonstrates that LLM-based measurement combined with structural modeling can capture core therapeutic processes at scale. Empathy and exploration showed small but consistent turn-level associations with client disclosure and emotional expression, whereas prior-session rapport did not show a robust association. These findings provide a computational foundation for precision modeling of psychotherapy processes and may offer actionable insights for scalable therapist training and AI-supported clinical education systems.
\end{abstract}

\keywords{psychotherapy process modeling; therapist empathy; client disclosure; structural equation modeling; large language models; automated behavioral coding; therapeutic alliance; digital mental health; AI-supported training}

\maketitle
\pagestyle{plain}

\section{Introduction}

Psychotherapy is an effective, evidence-based treatment for many common mental health conditions, yet access and quality remain persistent challenges. Digital mental health technologies---including teletherapy platforms \cite{robledoyamamoto2021teletherapy}, conversational agents \cite{vaidyam2019chatbots}, clinician-facing dashboards \cite{berardi2024barriers}, and training or supervision tools \cite{kenny2007virtual}---have expanded the reach of care and created new opportunities to monitor and improve treatment delivery. However, improved access alone does not guarantee effective care: clinical benefit depends heavily on \emph{how} therapeutic interactions unfold, including the clinician's moment-to-moment responses, the relational bond between therapist and client, and clients' in-session engagement and emotional processing \cite{eaton1988alliance,opland2024psychotherapy,dalfonso2020digital}.

A central limitation in both psychotherapy research and digital mental health implementation is that the mechanisms linking therapist behavior to patient outcomes are difficult to measure at scale. Therapeutic effectiveness is multifactorial, and the specific effects of discrete therapist behaviors on patient responses are often not characterized well \cite{hill2023psychotherapy}. In routine care and many digital health studies, evaluation relies on self-report symptom scales and post-session assessments of therapist-client interaction (e.g., PHQ-9, GAD-7, WAI, qualitative reflections) \cite{kroenke2001phq9,spitzer2006gad7,fossey2012selfrated,horvath1989wai}. While indispensable for capturing patient experience and clinical status, these measures are typically collected infrequently, provide limited resolution about what happens within sessions, can be influenced by recall and rater biases, and offer little actionable information for real-time clinical decision support or training.

A complementary approach is to quantify the therapy \emph{process} directly from session content. Prior work suggests that patient self-disclosure relates to anxiety and depression symptoms \cite{kim2021mediating}, that in-session emotional expression is associated with therapeutic change \cite{peluso2018emotional}, and that the therapeutic alliance---often conceptualized as bond (rapport), agreement on goals, and collaboration---is robustly linked to treatment success across modalities and populations \cite{eaton1988alliance,horvath1991alliance}. Yet, measuring these constructs at the granularity needed to support scalable evaluation and feedback remains challenging. Manual behavioral coding is labor-intensive and difficult to deploy in real-world digital care settings, motivating computational approaches that can extract clinically meaningful signals from therapy data at scale \cite{yang2024digital,sharma2020empathy,chikersal2020understanding,althoff2016counseling}.

In this study, we develop and evaluate a transcript-based computational framework for quantifying therapist--client dynamics in psychotherapy, with an emphasis on constructs that are clinically interpretable and potentially actionable in digital care contexts. We focus on three classes of variables: (1) \textbf{therapist behaviors} (empathic responses, reflective-listening, interpretations and exploratory queries), (2) \textbf{therapeutic relationship quality} (rapport as a bond-based indicator of therapeutic alliance), and (3) \textbf{client responses} (self-disclosure and emotional states). \textbf{Our objective is twofold:} first, to develop accurate and scalable measures of these constructs from real-world psychotherapy transcripts; and second, to empirically model how therapists' behaviors and relationship context are associated with clients' immediate responses during sessions.

To address measurement, we used large language models (LLMs) to score therapist empathy components (Interpretation, Exploration, Emotional Reaction, and Reflection \cite{sharma2020empathy}), therapist--client rapport \cite{leach2005rapport,elvins2008alliance}, and client responses including self-disclosure \cite{hill2019selfdisclosure} and emotions (anger, contempt, disgust, enjoyment, fear, sadness, surprise, anxiety, depression) \cite{ekman1999basic}. We applied these measures to psychotherapy transcripts from the Alexander Street corpus and validated automated measurements against ratings from trained human annotators. Across constructs, LLM-derived scores aligned well with human judgments (with ICC ranging from 0.60 to 0.81), supporting the feasibility of using LLMs as a scalable measurement layer for clinically grounded process analysis.

To model the associations between therapist behaviors and therapist-client rapport with client behaviors, we used Structural Equation Modeling (SEM) to test theory-motivated pathways linking therapist behavior, rapport, and client outcomes. Prior work has used SEM to examine causal and temporal relationships among communication variables and alliance-related constructs \cite{vail2022causal}. Building on this tradition, we evaluated whether therapists' immediate empathic and exploratory behaviors are associated with clients' subsequent disclosure and emotional expression, and whether cumulative rapport from prior sessions adds explanatory value beyond moment-to-moment behaviors. We tested the following four hypotheses: \emph{H1:} Therapist empathy and exploration (i.e., encouragement of deeper discussion) in the immediately preceding turn are associated with client self-disclosure. \emph{H2:} Cumulative rapport built prior to the current session is associated with client self-disclosure. \emph{H3:} Higher levels of therapist empathy and exploration are associated with greater expression of negative emotions. \emph{H4:} Higher levels of cumulative rapport between therapist and client are associated with greater expression of negative emotions. Our findings indicate that therapist empathy and exploration are associated with increased client disclosure and changes in emotional expression, whereas rapport in the prior session shows no association with client self-disclosure or negative emotion expression.

By combining validated, transcript-derived process measures with theory-driven modeling, this work contributes both measurement methods and empirical evidence that are directly relevant to digital medicine and healthcare delivery. The framework supports scalable assessment of therapeutic process variables that are difficult to capture with conventional symptom scales alone, and it enables empirical tests of hypotheses about how clinician behaviors relate to patient responses during care. These capabilities can inform digital mental health platforms, quality improvement initiatives, clinician training and supervision tools, and future clinician-facing decision support systems that require fine-grained, interpretable signals about therapeutic interactions.

\section{Methods}

\section{Dataset}

Alexander Street (AS) \cite{alexanderstreet2026} is a trusted source of psychological counseling materials for academics as it offers access to real clinical interactions that are typically hard to gather in large numbers due to practical limitations in clinical environments, including privacy concerns. The dataset comprises one-on-one therapy sessions transcribed by professional human transcribers who listened to the original audio recordings.

All transcripts were carefully anonymized following ethical research standards and HIPAA rules. Personal details were removed to protect client privacy while maintaining clinical relevance. Therapist names are unknown, but clients likely worked with the same therapist across sessions, while therapists probably worked with multiple clients.

We used strict selection criteria to maintain dataset consistency and meet our research goals. We kept only one-on-one counselor-client sessions and removed all group therapy, family therapy, and sessions with multiple participants. The final processed dataset includes 1,610 unique therapy sessions from 37 clients, covering 118 presenting conditions (e.g., anxiety, depression, burnout; clients could present with more than one condition) and a range of topics such as childhood trauma, sexual abuse, and social anxiety. The dataset comprises 243,407 utterances, counting both therapist and client turns, with a mean session length of 151.2 utterances (SD = 101.6). The analyses reported below model clients' behavior in a sample of 119,233 utterances from 1,573 sessions and 37 clients. We exclude the first session from each client because, by definition, it has no rapport value measured during a prior session. In this study, an \emph{utterance} refers to \textbf{a single speaking turn} in the therapy dialogue---an instance in which either the client or the therapist speaks before the conversational floor switches. The dataset also allows us to study concepts that can only be captured in a chunk of the conversation or across multiple turns, rather than at the single turn level.

\subsection{Ethical Considerations}

This paper used anonymized Alexander Street (AS) therapy transcripts under a non-exclusive license, in a manner consistent with U.S. fair use provisions and international law, and solely for research, educational, or other strictly non-commercial uses that do not involve public performance. All co-authors are authorized users of the dataset. Per NIH guidance, secondary research using data for which the research team has no access to identifiers does not meet the definition of human subjects research\footnote{\url{https://irbo.nih.gov/irb-review/not-human-subjects-research/}}. The transcripts were de-identified before release, and the research team had no access to identifiers or to any key linking transcripts to individuals. This study therefore does not constitute human subjects research, and IRB review was not required. The generative AI API we use to measure clinical behaviors does not use data submitted through the API to train or improve its models. Therefore, our use of generative AI models does not result in the AS dataset being used for model training.

\section{Measurements of Key Therapy-Related Constructs}

We first describe the constructs we measure (Constructs). We then describe the full development pipeline and validation studies (Prompt Design and Iteration; Validation with Human Annotators; Model Selection). The validation is conducted on a stratified subset of the AS dataset (see Stratified Sampling). The SEM modeling testing the relationships among rapport, therapist empathy, and client outcomes uses the full AS dataset.

\subsection{Constructs}

We automatically measure the following three types of constructs: \emph{client-side} (i.e., self-disclosure, emotions), \emph{therapist-side} (i.e., empathy skills), and the \emph{relationship} (i.e., rapport).

We required each construct to satisfy three conditions. First, it had to be operationalized by an existing validated framework, so that our prompts encode an established definition rather than a new one. Second, it had to be observable in transcribed text alone, since our corpus contains no audio or video. Third, it had to be measurable at the granularity our modeling requires, meaning turn level for therapist behaviors and client responses and session level for relational quality. Applying these criteria yields the constructs described below.

\subsubsection{Constructs about Therapist Behavior}

To ensure clinical validity and interpretability, each construct was grounded in established psychological theory and prior empirical frameworks. Because the dataset lacks reliable information about the treatment modalities or specific intervention techniques used by individual psychologists, we do not attempt to infer or evaluate modality-specific methods. Instead, we focus on cross-cutting psychotherapy ``common factors''---relational processes shared across approaches, such as empathy and rapport \cite{laska2014expanding,wampold2015common}. These theoretical and empirical foundations guided prompt design for each construct:

\begin{itemize}
\item
  \textbf{Empathy:} We model empathy through four interrelated components. The first three are drawn from the EPITOME framework \cite{sharma2020empathy}, which conceptualizes empathic communication as comprising: \emph{Emotional Reactions} (expressing emotions such as warmth, compassion, and concern), \emph{Interpretations} (communicating an understanding of the client's feelings and experiences), and \emph{Explorations} (attempting to understand the client's feelings and experiences through questions and probes). Each mechanism is scored on a 3-point scale (0 = not present, 1 = weak, 2 = strong), with total scores ranging from 0 to 6. The fourth dimension, \emph{Reflection}, follows definitions from Motivational Interviewing (MI) \cite{miller2023motivational,shah2022modeling,wu2022annomi} and is treated as a related but distinct construct that reinforces empathy. Reflection is defined as a therapist response that mirrors or builds upon the client's preceding statement. It is typically categorized as \emph{Simple} (paraphrasing client statements) or \emph{Complex} (inferring deeper meaning or emotion). Following prior computational work such as the PAIR framework \cite{min2022pair}, we adopt a binary Reflection vs. Non-Reflection scheme, using the Simple-Reflection/Complex-Reflection distinction as guiding examples within the prompt to calibrate model judgment and improve interpretability.
\end{itemize}

\subsubsection{Constructs about Therapist-Client Relationship}

\begin{itemize}
\item
  \textbf{Rapport:} The therapeutic alliance that the therapist and client develop over time is a robust predictor of improvements in clients' psychological wellbeing~\cite{laska2014expanding,wampold2015common}. Here, we focus on \emph{rapport}---the felt bond and harmonious responsiveness that undergirds alliance \cite{leach2005rapport}. Modeled as the ``Bond'' component of the therapeutic alliance, rapport reflects \emph{mutual liking}, \emph{confidence}, \emph{trust}, and \emph{affective attunement}. We adopt four items from the WAI-O-S bond subscale \cite{tichenor1989comparison} to measure rapport on a 7-point Likert scale. Because rapport is presumed to develop slowly over time, we measure it at the prior session rather than the prior utterance.
\end{itemize}

\subsubsection{Constructs about Client Behavior}

We measure two utterance-level components of client behavior: \emph{self-disclosure} and their \emph{emotions}. Together they characterize openness and affect as they unfold in session. \emph{Self-Disclosure} is the verbal revelation of personal information, experiences, or feelings. \emph{Emotion} is central to therapeutic interaction; clients are more likely to improve when they reveal their emotions~\cite{peluso2018emotional,greenberg2008emotion}.

\begin{itemize}
\item
  \textbf{Self-Disclosure:} Grounded in intimacy theory \cite{vondracek1971manipulation,bak2014selfdisclosure}, we classify utterances on a three-level scale: \emph{General} (G) for factual or impersonal content, \emph{Medium} (M) for personally relevant but non-sensitive expressions, and \emph{High} (H) for intimate or emotionally vulnerable disclosures \cite{balani2015detecting}.
\item
  \textbf{Emotion:} We assess emotional expression using nine sub-constructs. Our taxonomy integrates Ekman's seven universal emotions \cite{ekman1999basic}---\emph{enjoyment}, \emph{sadness}, \emph{anger}, \emph{disgust}, \emph{contempt}, \emph{fear}, and \emph{surprise}---augmented with two clinically relevant categories: \emph{anxiety and depression. The extent to which each of the nine emotions is expressed in each client utterance is rated on a 5-point Likert scale ranging from ``Very slightly or not at all'' to ``Extremely''.}
\end{itemize}

\subsection{Prompt Design and Iteration}

Each construct is operationalized through a dedicated LLM-based prompt aligned with validated psychological definitions (Table 1). All prompts share a common structure specifying the evaluator's role, output format, and evaluative constraints, while rubrics and rating scales are tailored to each construct (shown in Table 2). For maximal generalizability \cite{cheng2025revisiting}, we primarily employed zero-shot prompting, in which the prompts assume general knowledge and provide no examples. Rubrics were derived from clinical theory for all constructs (e.g., \cite{sharma2020empathy,miller2023motivational,tichenor1989comparison,vondracek1971manipulation,bak2014selfdisclosure}, Table 1), with inline examples added only for constructs requiring subtle distinctions \cite{hatanpaa2025distributional}. Following recent work on inference-time computation \cite{parashar2025inference,liu2025bag}, the empathy and self-disclosure prompts additionally use chain-of-thought reasoning, asking the model to justify its rating by citing specific elements of the utterance \cite{kojima2022large}. For each construct, the model first generated a rationale grounded in conversational evidence before producing the final rating. To limit run-to-run variability, every utterance was scored in three independent runs and the ratings averaged. Full prompt wordings appear in Appendix A.

We used construct-specific context windows, supplying only as much preceding dialogue as was needed to disambiguate relational or emotional cues and adapting window length to each construct's evaluative demands (details shown in Table 6 in Appendix).

We refined each construct prompt iteratively, evaluating variants against the annotated datasets summarized in Table 7 in Appendix B. Early exploration revealed consistent differences across foundation models in stability and sensitivity to prompt phrasing; GPT-4o-mini was selected as the primary model for its agreement with human annotators, its stability across prompt revisions, and the cost of annotating 243,407 counselor and client utterances; this initial choice was subsequently re-examined against five further models, as reported in Model Selection below. Two lessons recurred in the iterations: requiring the model to cite specific conversational cues improved scoring consistency, and simplifying definition phrasing reduced noise more effectively than elaborating it.

\begin{table}[tb]
\caption{This table provides a summary of the background in clinical theory.}
\label{tab:theory}
\small
\begin{tabularx}{\columnwidth}{@{}P{0.22\columnwidth}LL@{}}
\toprule
\textbf{Category} & \textbf{Constructs} & \textbf{Key Clinical Theory and Framework} \\
\midrule
Therapist-Side Behaviors & Empathy -- Emotional Reaction / Interpretation / Exploration & EPITOME Framework \cite{sharma2020empathy} \\
 & Empathy -- Reflection & Motivational Interviewing Criteria \cite{miller2023motivational} \\
\midrule
Client-Side Behaviors & Self-Disclosure & Intimacy Framework \cite{vondracek1971manipulation,bak2014selfdisclosure} \\
 & Emotion & Ekman’s Basic Emotions \cite{ekman1999basic}; PANAS Scale \cite{crawford2004panas} \\
\midrule
Therapeutic Relationship & Rapport & WAI-O-S Bond items \cite{tichenor1989comparison}: Items 3, 5, 7, 9 \\
\bottomrule
\end{tabularx}
\end{table}

\begin{table*}[tb]
\caption{This table presents the key components of each prompt, including the evaluator role, core definition phrase, and output format.}
\label{tab:prompts}
\small
\begin{tabularx}{\textwidth}{@{}P{0.17\textwidth}P{0.22\textwidth}LP{0.14\textwidth}@{}}
\toprule
\textbf{Construct} & \textbf{Assigned Role} & \textbf{Definition Snippet} & \textbf{Output Format} \\
\midrule
\textbf{Empathy} -- Emotional Reaction / Interpretation / Exploration & \emph{``...a professional evaluator specializing in assessing therapist empathy...''} & \emph{``...Emotional Reactions: Expressing warmth, compassion, or concern... Strong (2): Explicitly specifies emotions, clearly conveyed with warmth and understanding. Weak (1)...''} & 3 mechanisms (0--2 each) \\
\addlinespace
\textbf{Empathy} -- Reflection & \emph{``...a professional evaluator specializing in motivational interviewing...''} & \emph{``Reflection: A therapist's response that demonstrates understanding by restating or adding insight, or maintaining alignment with the client’s statements...''} & Binary: Reflection / Non-Reflection \\
\addlinespace
\textbf{Self-Disclosure} & \emph{``...a professional psychologist specializing in communication analysis and self-disclosure...''} & \emph{``M (Medium Disclosure): The client shares non-sensitive personal experiences, plans, or general life updates (e.g., `I started a new job last week...')''} & Ordinal: General / Medium / High \\
\addlinespace
\textbf{Emotion} & \emph{``...an expert conversation evaluator ... skilled in identifying subtle emotional cues ...''} & \emph{``Assess the central utterance using nine emotions. Rate each emotion on a 5-point Likert scale...''} & 9 emotions × 5-point Likert \\
\addlinespace
\textbf{Rapport} & \emph{``...an expert conversation evaluator assessing the therapeutic alliance...''} & \emph{``Assess bond quality across mutual liking, confidence, appreciation, and trust. Each rated on a 7-point Likert scale...''} & 4 aspects + overall (1--7) \\
\bottomrule
\end{tabularx}
\end{table*}

\subsection{Validation with Human Annotators}

To evaluate the effectiveness and reliability of our LLM-based annotation framework, we conducted human validation on a stratified sample of model-labeled utterances and segments.

\begin{itemize}
\item
  \textbf{Stratified Sampling} Our full dataset included 243,407 utterances drawn from 1,610 unique therapy sessions. Given this scale, we used stratified sampling strategies tailored to each construct to select representative and label-balanced subsets for validation.
\item
  \textbf{Recruitment \& Participation} Annotators were native English speaking university students who were either psychology majors or had completed at least two psychology- or counseling-related courses. Utterance-level classification tasks (self-disclosure and the empathy components) were each rated by three annotators; rapport and emotions, which involve longer context windows and multi-dimensional rating scales, were each rated by six. In total, \textbf{27} annotators took part, each contributing between 15 and 17 hours of annotation.
\item
  \textbf{Training} All annotators participated in structured training workshops tailored to each construct. Each session included definition reviews, guided examples, practice annotations, and collaborative discussions to ensure alignment in interpretation.
\end{itemize}

\subsection{Model Selection}

We applied identical prompts, with no model-specific tuning, to six models from three vendors: GPT-4o-mini, GPT-5.5, Claude Opus 4.8, Claude Sonnet 4.6, Gemini 3.5 Flash, and Gemini 3.1 Flash-Lite, scoring each against the same human-annotated validation sets. Exact version strings and API access dates are reported in Table 9 in Appendix B and pinned in our released analysis code.

\textbf{Agreement with human annotations was broadly consistent across models} (Table 8 in Appendix B): overall ICC ranged from 0.713 to 0.801 (SD 0.032), and every model exceeded 0.70 despite spanning roughly an order of magnitude in cost. We take this as evidence that measurement accuracy rests on the prompts and their grounding in established coding frameworks rather than on any particular vendor's model.

\textbf{The exception was emotion.} Under GPT-4o-mini, three of the nine emotion categories fell below ICC 0.50 (fear 0.453, anxiety 0.465, depression 0.489) and the mean emotion ICC was 0.606, which we did not regard as adequate for the outcome measures carrying H3 and H4. On the same validation sets the best-performing model reached a mean emotion ICC of 0.743, with every category above 0.65 (Table 8b). We therefore measured all therapist behaviors (empathy components, exploration, reflection), rapport, and client self-disclosure with GPT-4o-mini, and re-derived the nine emotion measures with Gemini 3.5 Flash, which achieved the highest overall agreement across construct families (mean ICC 0.787).

\begin{table*}[tb]
\caption{This table shows validation results comparing human--human reliability and Model--human agreement across constructs}
\label{tab:validation}
\begin{minipage}{\textwidth}\small
Note. Empathy, reflection, rapport, and self-disclosure values were produced with \textbf{GPT-4o-mini.} All nine emotion values were re-derived with \textbf{Gemini 3.5 Flash} following the comparative validation reported in Table 8; the corresponding GPT-4o-mini emotion values are reported in Table 8b. Because most constructs --- empathy mechanisms, self-disclosure, emotions, and rapport --- use ordinal or continuous scales, we report the intraclass correlation coefficient ICC(2,\emph{k}), a two-way random-effects coefficient that captures both absolute agreement and consistency in rating magnitudes \cite{koo2016guideline,mcgraw1996forming}. Reflection is binary; we report ICC(2,k) for it as well so that all constructs appear on a common scale, and its classification agreement is reported in Table 7.
\end{minipage}

\vspace{6pt}
\small
\begin{tabular}{@{}llll@{}}
\toprule
\textbf{Measurement} & \textbf{Human--Human ICC (2,k)} & \textbf{LLM--Human ICC (2,k)} & \textbf{LLM--Human Pearson} \(r\) \\
\midrule
\textbf{Empathy} & & & \\
\quad Emotional Reaction & \makecell[tl]{0.634 \\ \relax 95\% CI [0.59, 0.72]} & \makecell[tl]{0.601 \\ \relax 95\% CI [0.52, 0.67]} & 0.633 \\
\quad Interpretation & \makecell[tl]{0.763 \\ \relax 95\% CI [0.70, 0.82]} & \makecell[tl]{0.749 \\ \relax 95\% CI [0.69, 0.80]} & 0.814 \\
\quad Exploration & \makecell[tl]{0.826 \\ \relax 95\% CI [0.78, 0.87]} & \makecell[tl]{0.732 \\ \relax 95\% CI [0.67, 0.79]} & 0.686 \\
\quad Reflection & \makecell[tl]{0.654 \\ \relax 95\% CI [0.60, 0.74]} & \makecell[tl]{0.658 \\ \relax 95\% CI [0.60, 0.71]} & 0.754 \\
\addlinespace
\textbf{Rapport} & \makecell[tl]{0.720 \\ \relax 95\% CI [0.61, 0.81]} & \makecell[tl]{0.806 \\ \relax 95\% CI [0.75, 0.86]} & 0.676 \\
\addlinespace
\textbf{Self-Disclosure} & \makecell[tl]{0.749 \\ \relax 95\% CI [0.67, 0.81]} & \makecell[tl]{0.781 \\ \relax 95\% CI [0.70, 0.84]} & 0.621 \\
\addlinespace
\textbf{Emotions} & & & \\
\quad Anger & \makecell[tl]{0.799 \\ \relax 95\% CI [0.73, 0.85]} & \makecell[tl]{0.796 \\ \relax 95\% CI [0.745, 0.837]} & 0.679 \\
\quad Contempt & \makecell[tl]{0.657 \\ \relax 95\% CI [0.55, 0.73]} & \makecell[tl]{0.771 \\ \relax 95\% CI [0.718, 0.812]} & 0.710 \\
\quad Disgust & \makecell[tl]{0.719 \\ \relax 95\% CI [0.62, 0.79]} & \makecell[tl]{0.799 \\ \relax 95\% CI [0.744, 0.842]} & 0.711 \\
\quad Enjoyment & \makecell[tl]{0.838 \\ \relax 95\% CI [0.80, 0.87]} & \makecell[tl]{0.798 \\ \relax 95\% CI [0.732, 0.849]} & 0.673 \\
\quad Fear & \makecell[tl]{0.672 \\ \relax 95\% CI [0.56, 0.75]} & \makecell[tl]{0.706 \\ \relax 95\% CI [0.626, 0.768]} & 0.576 \\
\quad Sadness & \makecell[tl]{0.719 \\ \relax 95\% CI [0.60, 0.79]} & \makecell[tl]{0.733 \\ \relax 95\% CI [0.676, 0.782]} & 0.615 \\
\quad Surprise & \makecell[tl]{0.702 \\ \relax 95\% CI [0.61, 0.77]} & \makecell[tl]{0.677 \\ \relax 95\% CI [0.583, 0.749]} & 0.526 \\
\quad Anxiety & \makecell[tl]{0.757 \\ \relax 95\% CI [0.67, 0.81]} & \makecell[tl]{0.748 \\ \relax 95\% CI [0.695, 0.792]} & 0.629 \\
\quad Depression & \makecell[tl]{0.710 \\ \relax 95\% CI [0.62, 0.77]} & \makecell[tl]{0.659 \\ \relax 95\% CI [0.591, 0.719]} & 0.602 \\
\bottomrule
\end{tabular}
\end{table*}

\subsection{Interpretation of the Validation Results}

Table 3 shows that, according to established guidelines \cite{cicchetti1994guidelines}, all of our human--human and LLM--human ICC values fall in the ``good'' (0.60--0.74) to ``excellent'' (0.75--1.00) ranges.

Each of the four \textbf{\emph{Empathy}} dimensions was validated on 50 utterances by three raters, and all fell in the ``good'' to ``excellent'' range. Agreement was highest for Interpretation (ICC = 0.749) and Exploration (ICC = 0.732), and lower for Emotional Reactions (ICC = 0.601) and Reflection (ICC = 0.658) --- the same two dimensions that showed the lowest human--human reliability, suggesting they involve more subjective affective inference that is harder for both humans and LLMs to judge consistently.

\emph{\textbf{Self-Disclosure}} (150 utterances, three raters) and \emph{\textbf{Rapport}}, measured at the segment level (200 samples, six raters), both showed good to excellent agreement (ICC = 0.781 and 0.806, respectively), indicating that the model captures both the depth of client expression and multi-turn relational signals.

\textbf{\emph{Emotion}} was measured with Gemini 3.5 Flash and validated on 450 utterances spanning nine affect categories, annotated by six raters. Agreement averaged ICC 0.743 across the nine categories. Emotions with distinct verbal signatures were rated most reliably --- disgust (0.799), enjoyment (0.798), anger (0.796), and contempt (0.771) --- while more introspective or ambiguous states proved harder: anxiety (0.748) and sadness (0.733) still reached good agreement, but fear (0.706), surprise (0.677), and depression (0.659) were the weakest in our framework, reflecting the difficulty of inferring complex affect from text alone, without vocal tone or facial expression.

Overall, agreement with trained lay annotators ranged from good to excellent. Our annotators were undergraduate psychology majors or students who had completed at least two relevant courses; they were trained on our coding scheme but are not clinicians, so these figures should not be read as agreement with expert clinical judgment. We also note that an ICC above 0.75 is the conventional standard for measures intended to inform clinical decisions, and that our measures do not uniformly meet it.

\section{Modeling of Therapist--Client Dynamics in Psychotherapy}

We employed \textbf{Structural Equation Modeling (SEM)} to analyze the Alexander Street data in order to capture temporal patterns unfolding between therapist and client behaviors across sessions. To reduce model complexity, we first conducted \textbf{Principal Components Analysis (PCA)} for constructs such as emotion and empathy, which were composed of multiple components \cite{jolliffe2002pca}. We then specified and fit SEMs using R's lavaan package \cite{rosseel2012lavaan}. Because utterances are nested within sessions and sessions within clients, we estimated a saturated multivariate SEM with standard errors clustered on a unique session identifier constructed as the combination of client and session \cite{kline2016principles}. To verify that our conclusions did not depend on this choice of statistical method, we also fitted a multilevel model with random intercepts for session and for client. The two specifications agree on the direction and approximate magnitude of every association; the only discrepancy concerns rapport, whose association with self-directed negative emotion is marginal in the SEM and marginally significant in the multilevel model, which is the basis for our statement that the rapport effect is not robust.

\subsection{Results}

\subsubsection{Principal Components Analysis (PCA) Results}

\paragraph{Emotions}

Across nine client emotion indicators (\(N = 122{,}939\)), PCA produced a four-component rotated structure explaining 78.9\% of the total variance. The first component, \emph{Self-Directed Negative Emotions}, showed strong positive associations with \emph{sadness} (.88), \emph{anxiety} (.85), \emph{fear} (.74), and \emph{depression} (.88). This component captures inwardly oriented emotional distress; averaging these four emotions leads to a highly reliable \emph{Self-Directed Negative Emotions} scale (Cronbach Alpha = .87) . The second component, \emph{Outward-Directed Negative Emotions}, was defined by high associations with \emph{contempt} (.91), \emph{disgust} (.79), and \emph{anger} (.79), representing externally oriented aversive reactions directed towards some external target (e.g., angry at \ldots). Averaging these three emotions leads to a highly reliable \emph{Outward-Directed Negative Emotions} scale (Cronbach α = .79). The third component corresponded almost exclusively to \emph{surprise}, which showed minimal cross-loadings, indicating that surprise functions as an independent expressive signal rather than belonging to either negative-emotion grouping. Similarly, the fourth component corresponded almost exclusively to \emph{enjoyment}, which showed an extremely high loading (.98) and low cross-loadings. We treat these four components as distinct outcomes: self-directed negative emotions, outward-directed negative emotions, surprise, and enjoyment, each of which functions as an independent expressive signal.

\paragraph{Empathy}

For the four therapist-behavior items (\(N = 120{,}277\)), PCA revealed a two-component solution explaining roughly 75\% of the variance. The first component, \emph{General Empathy}, showed strong associations with \emph{interpretation} (.62), \emph{reflection} (.57), and \emph{reaction} (.49), representing empathic understanding and responsive attunement. Averaging these resulted in a highly reliable General Empathy scale (Cronbach α = .71). The second component, \emph{Exploration}, emerged as a distinct factor defined primarily by the \emph{exploration} item (.94), indicating a separate behavioral dimension focused on directive inquiry and probing.

\begin{table*}[tb]
\caption{PCA Loadings of Emotion and Empathy}
\label{tab:pca}
\small
\begin{tabular}{@{}lllll@{}}
\toprule
Item & Self-Directed Negative Emotions & Outward-Directed Negative Emotions & Surprise & Enjoyment \\
\midrule
\textbf{Panel A. Emotions} & & & & \\
Anger & 0.236 & \textbf{0.788} & 0.066 & -0.098 \\
Contempt & 0.038 & \textbf{0.905} & 0.026 & -0.021 \\
Disgust & 0.062 & \textbf{0.788} & 0.049 & 0.003 \\
Enjoyment & -0.137 & -0.045 & 0.061 & \textbf{0.984} \\
Fear & \textbf{0.744} & -0.056 & 0.279 & -0.101 \\
Sadness & \textbf{0.879} & 0.167 & -0.136 & -0.077 \\
Surprise & -0.007 & 0.075 & \textbf{0.929} & 0.076 \\
Anxiety & \textbf{0.848} & 0.065 & 0.218 & -0.147 \\
Depression & \textbf{0.880} & 0.129 & -0.173 & -0.073 \\
\midrule
\textbf{Panel B. Empathy} & & & & \\
Reaction & \textbf{0.494} & −0.331 & & \\
Interpretation & \textbf{0.624} & 0.002 & & \\
Exploration & 0.209 & \textbf{0.942} & & \\
Reflection & \textbf{0.568} & −0.060 & & \\
\bottomrule
\end{tabular}

\vspace{4pt}
\begin{minipage}{\textwidth}\small
Note. Panel A reports varimax-rotated loadings from the four-component solution on the Gemini 3.5 Flash emotion measures (N=122,939), which together explain 78.9\% of the variance. Panel B reports the therapist-behavior loadings (N=120,277), which together explain 75\% of the variance. Both PCA sample sizes exceed the estimation sample of 119,233 because they include utterances from each client's first session, which the models drop for lack of a prior-session rapport value; the empathy solution is estimated on therapist utterances rather than client utterances.
\end{minipage}
\end{table*}

\subsubsection{Structural Equation Modeling (SEM) Modeling Results}

Based on these PCA results, we specified a multivariate SEM in which five utterance-level client-related outcome variables (disclosure and four emotional expression variables -- self-directed negative emotion, outward-directed negative emotion, surprise, and enjoyment) were regressed on utterance-level therapist behaviors (exploration, empathy), session-level rapport from the previous session, and the number of sessions clients have had with their therapist (log-transformed session index).

We estimated a structural equation model in which the five client outcomes in a given utterance were regressed simultaneously on the therapist's behaviors in the preceding utterance, client--therapist rapport from the previous session, and the log-transformed session index, with standard errors clustered on a unique session identifier constructed as the combination of client and session. Because every outcome is regressed on every predictor and all residual covariances are free, the model is saturated and has zero degrees of freedom; fit indices are consequently uninformative and are not reported. The SEM estimates are shown in Table 5.

As a robustness check, we conducted parallel analyses using multilevel models. Utterances are nested within sessions and sessions within clients, and we therefore estimated random intercepts for both session and client. Results are reported in Table 12 in the appendix and are consistent with the SEM estimates. Because these models are not saturated, likelihood-based fit comparisons are informative, and we additionally report marginal and conditional R² for each outcome (Tables 13 and 14).

\begin{table*}[tb]
\caption{This table shows the Structural Equation Model (SEM) results.}
\label{tab:sem}
\begin{minipage}{\textwidth}\small
Note. N=119,233 client utterances from 1,573 sessions and 37 clients, with standard errors clustered on a unique session identifier constructed as the combination of client and session. Values rounded to three decimals. Estimates in bold have P\textless.05; exact P values and 95\% confidence intervals are given in the table.
\end{minipage}

\vspace{6pt}
\small
\begin{tabular}{@{}P{0.15\textwidth}P{0.11\textwidth}P{0.17\textwidth}P{0.17\textwidth}P{0.11\textwidth}P{0.11\textwidth}@{}}
\toprule
\textbf{Predictor} & \textbf{Client’s disclosure} & \multicolumn{4}{l}{\textbf{Client’s emotion expression}} \\
\cmidrule(l){3-6}
 & & \textbf{Self-directed negative emotion} (Sadness, Fear, Anxiety, Depression) & \textbf{Outward-directed negative emotion} (Anger, Contempt, Disgust) & \textbf{Surprise} & \textbf{Enjoyment} \\
\midrule
\textbf{log Session ID} & \makecell[tl]{0.008 \\ \relax [−0.009, 0.025] \\ \relax \emph{P}=.35} & \makecell[tl]{0.002 \\ \relax [−0.020, 0.024] \\ \relax \emph{P}=.85} & \makecell[tl]{0.019 \\ \relax [−0.001, 0.040] \\ \relax \emph{P}=.06} & \makecell[tl]{0.011 \\ \relax [−0.001, 0.023] \\ \relax \emph{P}=.07} & \makecell[tl]{0.008 \\ \relax [−0.011, 0.028] \\ \relax P=.40} \\
\addlinespace
\textbf{Rapport (prior session)} & \makecell[tl]{−0.003 \\ \relax [−0.017, 0.012] \\ \relax \emph{P}=.73} & \makecell[tl]{−0.017 \\ \relax [−0.035, 0.001] \\ \relax \emph{P}=.07} & \makecell[tl]{−0.004 \\ \relax [−0.020, 0.012] \\ \relax \emph{P}=.62} & \makecell[tl]{0.002 \\ \relax [−0.008, 0.012] \\ \relax \emph{P}=.66} & \makecell[tl]{\textbf{0.017} \\ \relax [0.001, 0.033] \\ \relax \emph{P}=.04} \\
\addlinespace
\textbf{Exploration} & \makecell[tl]{\textbf{0.086} \\ \relax [0.078, 0.094] \\ \relax \emph{P}<.001} & \makecell[tl]{\textbf{0.078} \\ \relax [0.070, 0.087] \\ \relax \emph{P}<.001} & \makecell[tl]{\textbf{−0.011} \\ \relax [−0.017, −0.004] \\ \relax \emph{P}=.002} & \makecell[tl]{\textbf{−0.013} \\ \relax [−0.019, −0.007] \\ \relax \emph{P}<.001} & \makecell[tl]{\textbf{−0.041} \\ \relax [−0.048, −0.034] \\ \relax \emph{P}<.001} \\
\addlinespace
\textbf{Empathy} & \makecell[tl]{\textbf{0.086} \\ \relax [0.078, 0.095] \\ \relax \emph{P}<.001} & \makecell[tl]{\textbf{0.118} \\ \relax [0.110, 0.127] \\ \relax \emph{P}<.001} & \makecell[tl]{\textbf{0.020} \\ \relax [0.013, 0.028] \\ \relax \emph{P}<.001} & \makecell[tl]{\textbf{0.010} \\ \relax [0.004, 0.017] \\ \relax \emph{P}=.002} & \makecell[tl]{\textbf{−0.037} \\ \relax [−0.043, −0.030] \\ \relax \emph{P}<.001} \\
\bottomrule
\end{tabular}
\end{table*}

\begin{figure*}[tb]
\centering
\includegraphics[width=0.64\textwidth]{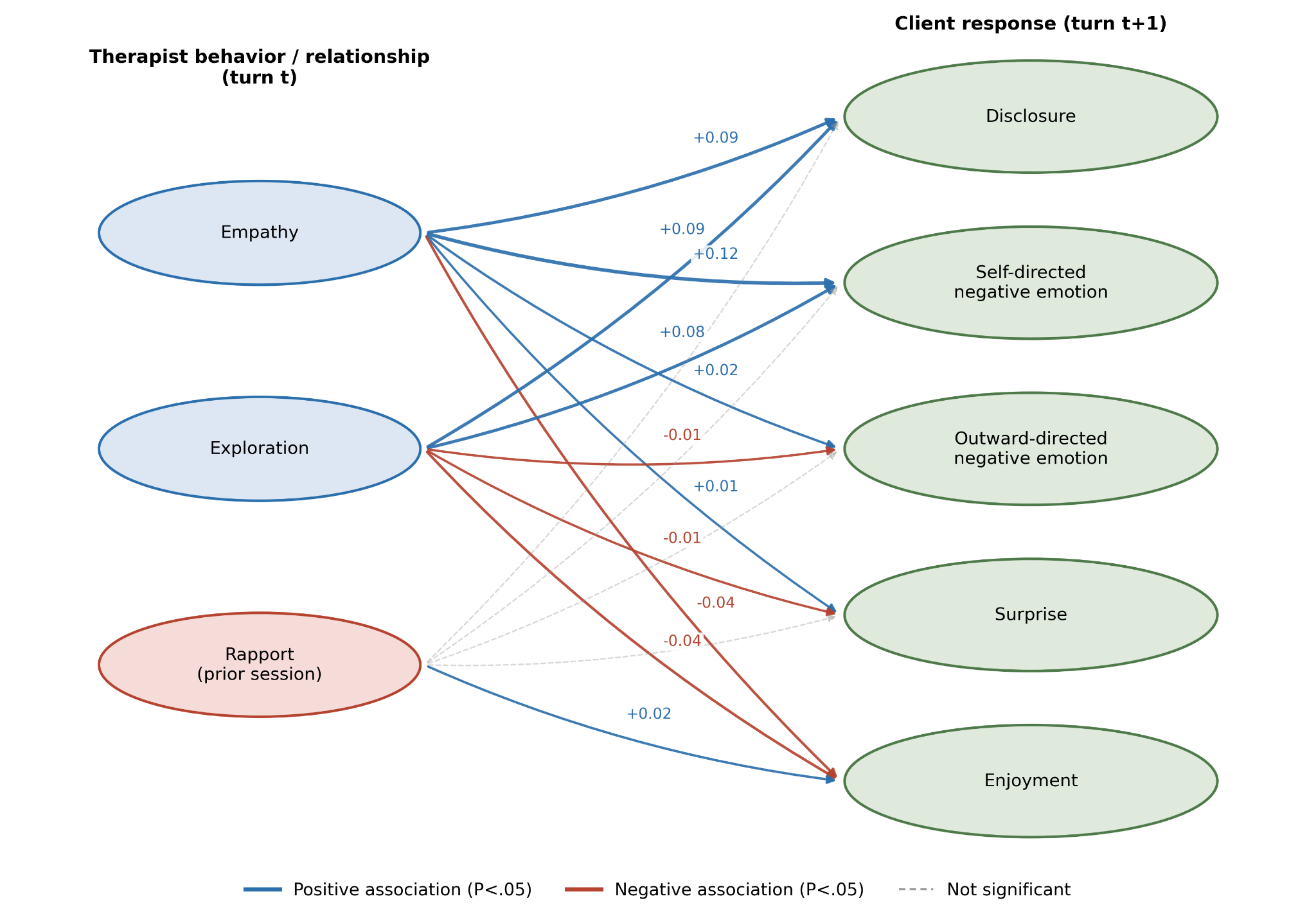}
\caption{Path diagram of associations between therapist behavior, prior-session rapport, and client responses in the following turn. Estimates are standardized coefficients from the saturated SEM reported in Table 5, with standard errors clustered at the session level (N=119,233). Line thickness is proportional to effect magnitude. Solid blue indicates a significant positive association and solid red a negative one at P\textless.05; dashed grey indicates a non-significant association. Confidence intervals are reported in Table 5. The log-transformed number of sessions the client already had was included as a covariate in SEM analysis reported in Table 5, but omitted in the path diagram. All associations are small in absolute terms (\textbar β\textbar{} ≤ 0.12).}
\Description{Path diagram linking empathy, exploration, and prior-session rapport (left) to disclosure, self-directed negative emotion, outward-directed negative emotion, surprise, and enjoyment (right), with arrows labeled by standardized coefficients.}
\label{fig:path}
\end{figure*}

\begin{enumerate}
\item
  \textbf{Hypothesis 1 is supported: A client's self-disclosure is associated with the counselor's immediately preceding behavior, particularly their expressed empathy and exploratory probing.}

  \begin{itemize}
\item
  As shown in Table 5, therapist behaviors in the preceding turn were associated with client self-disclosure. Both empathy (β=0.086, 95\% CI 0.078 to 0.095, P\textless.001) and exploration (β=0.086, 95\% CI 0.078 to 0.094, P\textless.001) showed small positive associations. The two coefficients were nearly identical in magnitude, indicating that expressed empathy and exploratory questioning in the prior therapist turn were similarly associated with clients' likelihood of disclosing personal information in the subsequent turn. These findings support Hypothesis 1.
\end{itemize}
\item
  \textbf{Hypothesis 2 is not supported: A client's self-disclosure during a session is unrelated to cumulative rapport prior to the current session.}

  \begin{itemize}
\item
  \emph{Rapport} from the previous session showed no meaningful association with disclosure in the current session (β=−0.003, 95\% CI −0.017 to 0.012, P=.73). The effect was essentially zero and statistically nonsignificant, providing no evidence that cumulative rapport increases the amount of information clients choose to reveal. Thus, Hypothesis~2 is not supported.
\end{itemize}
\item
  \textbf{Hypothesis 3 is partially supported: Higher levels of therapist empathy and exploration are associated with greater expression of self-directed negative emotions.}

  \begin{itemize}
\item
  Therapist \emph{empathy} in the prior turn was positively associated with both types of negative emotion, with the association for self-directed negative emotion (β=0.118, 95\% CI 0.110 to 0.127, P\textless.001) approximately six times larger than for outward-directed negative emotion (β=0.020, 95\% CI 0.013 to 0.028, P\textless.001):

  \begin{itemize}
  \item
    \emph{Self-Directed Negative Emotions}: β=0.118, 95\% CI 0.110 to 0.127, P\textless.001
  \item
    \emph{Outward-Directed Negative Emotions}: β=0.020, 95\% CI 0.013 to 0.028, P\textless.001
  \item
    Exploration showed the same pattern for self-directed negative emotion (β=0.078, 95\% CI 0.070 to 0.087, P\textless.001) but was associated with slightly lower outward-directed negative emotion (β=−0.011, 95\% CI −0.017 to −0.004, P=.002). Hypothesis 3 therefore holds for self-directed negative emotion but not uniformly across both emotion groups.
  \end{itemize}

  This asymmetry aligns with the functional distinction between the two emotion groups. Empathic responses may accompany a relational context in which clients feel safer acknowledging internal suffering, which would be consistent with the observed pattern of greater sadness, anxiety, and depression alongside less enjoyment. Outward-directed emotions such as anger, contempt, and disgust, by contrast, are more often reactive to external circumstances and may be relatively less dependent on empathic attunement for their expression.
\end{itemize}
\item
  \textbf{Hypothesis 4 is not supported: Higher rapport between therapists and clients is not associated with greater expression of negative emotions.}

  \begin{itemize}
\item
  Contrary to Hypothesis 4, greater rapport in the previous session was associated with lower levels of self-directed negative emotion in the current session, although this association was not significant (β=−0.017, 95\% CI −0.035 to 0.001, P=.07). The association with outward-directed emotion was also nonsignificant:

  \begin{itemize}
  \item
    \emph{Self-Directed Negative Emotions}: β=−0.017, 95\% CI −0.035 to 0.001, P=.07
  \item
    \emph{Outward-Directed Negative Emotions}: β=−0.004, 95\% CI −0.020 to 0.012, P=.62
  \end{itemize}

  We found no evidence that higher rapport is associated with greater expression of internally directed negative emotions (sadness, fear, anxiety, depression) or outward-directed negative emotions (anger, contempt, disgust). However, higher rapport is associated with greater enjoyment (β=0.017, 95\% CI 0.001 to 0.033, P=.04).
\end{itemize}
\end{enumerate}

\section{Discussion}

\subsection{Summary of Findings and Interpretations}

This study examined how therapist behaviors and prior-session rapport relate to what clients say and feel during therapy. By analyzing each utterance and separating negative emotions into two groups, self-directed and outward-directed, we were able to see how different therapist actions are associated with different types of emotional expression and client disclosure.

We found that therapist empathy and exploration showed small positive associations with client disclosure (β=0.086 for each). When therapists responded with understanding or asked clients to elaborate, clients became more willing to share personal information in the next turn. Empathy was also associated with greater negative emotional expression, although the association was approximately six times larger for self-directed negative emotions (sadness, fear, anxiety, depression) than for outward-directed emotions (anger, contempt, disgust). Overall, these findings are consistent with literature reporting that empathy is associated with greater emotional accessibility and tolerance, which may help clients approach and articulate difficult feelings \cite{pascualleone2007emotional,elliott2018empathy}.

Contrary to our expectations, rapport was not associated with greater self-disclosure or outward-directed negative emotions. A methodological gap may help clarify the mixed findings for rapport. Our hypotheses assumed that rapport represented the client's subjective experience of trust and safety, but in this study, the LLM measurement of rapport was based on observer ratings (WAI-O). While prior research shows that both the outside observer and client version of the working alliance inventory predict positive treatment outcomes, like improvements in depression, anxiety, and post-traumatic stress, the effect sizes are larger for alliance measured from a client's perspective \cite{horvath1991alliance,fluckiger2018alliance}. However, the client and observer perspectives are not strongly related to each other. If the rapport scores reflect what observers see in the session, rather than how clients personally experience the relationship, then the weaker or unexpected effects become easier to understand. Future work should include measures based on client-rated rapport to better test our hypotheses.

\subsection{Limitations}

\subsubsection{Small Effect Size}

We acknowledge the small effect sizes of all the associations we report and that practical use of our empirical research would require evidence of stronger association. The weak effect size problem may be methodological and may require better automated measures of the constructs like empathy and self-disclosure. Or the low associations may be more theoretical, e.g., identifying the aggregation of turns and lags to show stronger associations. For example, our measure of rapport looks only at rapport as measured in the last session, but meta-analytic work \cite{fluckiger2018alliance} suggests measuring rapport over a longer span and measured later in therapy. Similarly, the relationship between empathy and self-disclosure may be stronger if empathy were measured over a longer period of time and not just at the immediately preceding turn.

\subsubsection{Dataset Scope and Generalizability}

Our analyses draw on a single corpus of professionally transcribed, one-on-one therapy sessions from the Alexander Street dataset. While this corpus provides high-quality, naturalistic clinical interactions, it reflects primarily Western therapeutic contexts. Linguistic, cultural, and religious backgrounds play an important role in shaping how clients express emotion, disclose personal information, and build rapport. Cultural norms influence preferred communication styles, expectations of therapist--client roles, and the degree to which emotional experiences are expressed directly or indirectly \cite{sue2016counseling,markus1991culture}. In many cultural contexts, emotional expression is modulated by sociocultural rules (e.g., display norms), and self-disclosure is shaped by collective versus individualistic value systems and culturally grounded understandings of distress \cite{kleinman1988rethinking}. These differences can affect how therapeutic processes unfold and how constructs such as empathy, disclosure, or rapport manifest in session dialogue. In addition, this dataset is text-only, focusing on verbal content rather than multimodal cues. Because many therapeutic constructs are conveyed through language, the corpus supports meaningful measurement of therapist behaviors and client processes, though future work incorporating audio or multimodal data could further enrich assessments of emotional nuance and relational attunement. Therefore, the generalizability of our findings to other modalities, populations, or cultural settings remains an open question.

We were also unable to conduct subgroup analyses by clinical status or participant characteristics, which could have revealed meaningful heterogeneity. Three features of the data prevent them. The corpus is anonymized, so patient names, gender, and other identifying details are unavailable. The recorded presenting conditions are vague and are associated at the client level, leaving roughly one to two clients per condition, far too few to estimate a separate model per subgroup. Most fundamentally, all measures in our models are standardized within client, which removes between-client variance by construction, so the model as specified cannot address between-client questions of any kind. Datasets that retain clinical and demographic metadata would permit the analyses this limitation forecloses.

\subsubsection{Contributions and Limitations of the LLM Prompts}

Table 8 shows that our prompts have the potential to generalize and are not an artifact of one vendor's model. If a specific model is updated or withdrawn, the prompts can be re-applied to an alternative model and yield comparable measurements. Researchers can also consult Table 8 to choose a model appropriate to their budget and construct of interest. However, we want to acknowledge three limitations:

\textbf{Portability of the prompts has limits.} For example, the ICCs for the emotion measures ranged from 0.606 to 0.743. For these constructs, model choice materially affects measurement, and researchers substituting one model for another should expect different results.

\textbf{Vendor-driven behavioral drift.} We recorded the exact model version strings and API access dates for all six models, pinned those versions in the released analysis code, and released all prompts. However, vendor-driven behavioral drift---the changes in behavior caused by shifts in the underlying models---is a limitation that we want to acknowledge \cite{chen2024chatgpt}.

\textbf{Lack of transparency.} Although the scoring prompts/rubrics can be directly inspected and critiqued and are released with the paper, the model's reasoning remains opaque \cite{ghassemi2021false}.

\subsubsection{Ethical Considerations in LLM-Based Assessments for Therapeutic Applications}

The use of LLMs to assess therapeutic interactions introduces important ethical considerations that must be carefully addressed to prevent harm and ensure responsible deployment. Although this study focuses on designing LLM-based measurement of generic therapist behavior types (e.g., empathy) and high-level client outcomes (e.g., self-disclosure and emotional expression), we also acknowledge that LLMs may inadvertently reproduce or amplify harmful behaviors, particularly those affecting marginalized or underrepresented groups. In the context of our study, these biases may manifest in the inappropriate or inaccurate evaluation of therapist behaviors or their relationships with client outcomes due to a limited understanding of the complex and culturally situated needs for different demographic groups. For example, models may systematically misinterpret communication styles common in particular cultures as avoidance, or misinterpret culturally normative emotional restraint as disengagement \cite{eng2012emotion,lim2016cultural}. Similarly, expressions of emotional state and symptoms from marginalized groups may be flattened or pathologized as has been the case across history when it comes to cultural differences for mental health treatment outcomes \cite{jimenez2022centering,pederson2023management}. Limitations of the current model include insufficient cultural responsiveness in non-Western therapeutic dyads \cite{yang2026exploring}, a lack of nuanced understanding of clinical context (e.g., the presence of underlying psychotic processes or approach-avoidance patient behaviors driven by insecure attachment), and a lack of external validation demonstrating clear value for patient outcomes.

\subsection{Future Work}

These considerations highlight several directions for future research. First, future studies should examine sessions conducted by therapists with varying levels of experience, as expertise may influence the variability of rapport. Second, future work should test sequential mediation and moderation directly. Our data could not test moderation: the rapport main effect was not robust across specifications, and 37 clients is too few to estimate a cross-level interaction with useful precision. Future work should employ multilevel dynamic SEM or related state-space/continuous-time approaches with distributed lags, random slopes, and cross-level interactions to capture heterogeneity across therapists and clients. Future work with datasets that include therapist identifiers could extend this work by examining how individual therapists differentially enact these general skills, but this lies beyond the scope of our present aims. Future analyses could also look for nonlinear patterns (such as thresholds or inverted-U shapes) and use mixture models to uncover different types of process patterns that might be hidden when averaging across all sessions.

Beyond these methodological directions, two ethical requirements should guide any future use of LLM-based automatic assessment. First, external validation before deployment. Our measures have been validated against trained lay annotators, not against clinical experts or patient outcomes. Demonstrating that these constructs predict clinically meaningful change is a prerequisite for any decision-support role. Until such validation exists, the appropriate use of this framework is research at scale, not clinical guidance.

Second, requirements for prospective deployment. Should systems of this kind eventually offer assessment or direct feedback to clinicians in training, that step would require safeguards our archival study does not need and therefore does not test: post-deployment monitoring for drift in the underlying commercial models, crisis-escalation protocols for sessions where risk indicators appear, rigorous data accountability for highly sensitive session content, and evaluation of differential performance across the populations on which the system would be used.

\section{Conclusion and Broader Implications}

This study provides a process-level account of how clinicians' in-session behaviors relate to clients' next-turn responses. Across outcomes, therapist empathy and exploration showed small but consistent associations with client disclosure and with the direction of emotional expression, while prior-session rapport showed no association that was robust across model specifications. Whether rapport conditions the association between therapist behavior and client response is a question our design cannot answer, and we identify it as a priority for future work. These findings provide a computational foundation for precision modeling of psychotherapy processes and may offer actionable insights for scalable therapist training and AI-supported clinical education systems.

\section{Disclosure of Delegation to Generative AI}

The authors declare the use of generative AI (GAI) in the research and writing process. According to the GAIDeT taxonomy (2026), the following tasks were delegated to GAI tools under full human supervision: (1) Reproducibility testing, (2) Proofreading and editing, (3) Reformatting. Responsibility for the final manuscript lies entirely with the authors. GAI tools are not listed as authors and do not bear responsibility for the final outcomes. Using LLMs as measurement instruments is described in Methods and is separate from this disclosure.

\section*{Data Availability}

The Alexander Street database \cite{alexanderstreet2026} is available through academic, public, and K--12 institutional library subscriptions. The measurement prompts and all information necessary to reproduce the analysis is provided in the supplementary materials.

\section*{Author Contribution}

AC: Conceptualization, Data curation, Methodology, Formal analysis, Writing -- original draft. SJ, CW: Data curation, Methodology, Validation. HZ: Conceptualization, Methodology, Formal analysis, Funding acquisition, Investigation, Supervision, Writing -- original draft, review \& editing. REK: Conceptualization, Formal analysis, Funding acquisition, Writing -- review \& editing. HS, TW: Conceptualization, Funding acquisition, Writing -- review \& editing.

\section*{Conflicts of Interest}

The authors do not serve on this journal's editorial board. No authors have acted as expert witnesses in relevant legal proceedings or have served or currently serve on committees of organizations that may benefit from the publication of this paper. Neither the authors nor their family members work for or own stocks or stock options in any company that developed or owns an app or product evaluated or discussed in this paper, nor do they otherwise stand to benefit financially from the publication of this article.

\section*{Funding Statement}

This work is funded by the National Institute of Mental Health (NIMH), ``SCH: Training Mental Health Supporters with Virtual Patients and Automated Feedback'' (R01 MH139114), and a fellowship provided by the Center for Machine Learning and Health at Carnegie Mellon University.

\bibliographystyle{ACM-Reference-Format}
\bibliography{references}

\appendix

\section*{Appendix A: Constructs Prompts}

\subsection*{Empathy - EPITOME}

\begin{lstlisting}
"You are a professional evaluator specializing in assessing therapist empathy. Your evaluations are based on clinical best practices, focusing on accuracy, nuance, and consistency."

"""
Task: Evaluate the therapist's response for empathy based on the client's post. Assign ratings for 3 empathy mechanisms Emotional Reactions, Interpretations, and Explorations based on their definitions and criteria.

Empathy Definitions:
Empathy Mechanism Definitions & Rating Criteria:
- Emotional Reactions: Expressing warmth, compassion, or concern.
  - Strong (2): Explicitly specifies emotions.
  - Weak (1): Alludes to emotions without explicitly labeling them. 
  - None (0): Doesn't convey any emotional reaction.
Note: Consider responses that express empathy indirectly or in a passive tone.

- Interpretations: Showing cognitive understanding of the client's feelings.
  - Strong (2): Specifies the inferred feelings or experiences or relates to similar personal or shared experiences that demonstrate understanding.
  - Weak (1): Acknowledges understanding but lacks specificity.
  - None (0): Doesn't demonstrate any understanding of client's emotions or experiences.

- Explorations: Encouraging deeper discussion of unstated client experiences.
  - Strong (2): Asks specific questions that label the client's experience.
  - Weak (1): Uses generic questions.
  - None (0): Doesn't include exploratory elements.

Instructions:
- Use the client's speech as context for evaluating the therapist's response.
- Assign ratings (0: None, 1: Weak, 2: Strong) for each mechanism based on the empathy mechanism definitions.
- Provide a brief explanation citing specific phrases from the therapist's response.

Conversation:
Client's Speech: "{client_speech}"  
Therapist's Response: "{therapist_response}"

Evaluation Output Format:
- Explanation (<= 70 words): [justify your rating, citing specific elements of the utterance if applicable.]
- Emotional Reactions: [0/1/2] 
- Interpretations: [0/1/2]
- Explorations: [0/1/2]
"""
\end{lstlisting}

\subsection*{Empathy - Reflection}

\begin{lstlisting}
"You are a professional evaluator specializing in motivational interviewing and therapeutic communication analysis. You must apply consistent, evidence-based assessment to determine whether the therapist's response qualifies as a 'Reflection' or 'Non-Reflection'."

"""
Task: Evaluate whether the therapist's utterance qualifies as a "Reflection" or "Non-Reflection" based on the provided conversation context and definitions.

Definitions:
- Reflection: A therapist response that demonstrates understanding of the client's statements by restating content, adding insight, or maintaining alignment with the client's statements:
    - Simple Reflection: A response that restates or slightly rephrases the client's statement without adding significant new meaning.
    - Complex Reflection: A response that expands on the client's statement by adding depth, insight, or new meaning.
- Non-Reflection: Any response that does not meet the criteria for a Reflection.

Instruction:
- Use the client's utterance as context for evaluating the therapist's response.
- Evaluate the therapist's response to determine if it aligns with the criteria of Simple Reflection, Complex Reflection, or Non-Reflection.
- Provide a brief explanation, citing specific phrases from the therapist's response if applicable.
    
Conversation:
Client's Post: "{client_utt}"
Therapist's Response: "{therapist_utt}"

Output Format:
Reasoning (<= 70 words): [justify your classification, citing specific elements of the utterance if applicable.]
Label: [Reflection/Non-Reflection]
"""
\end{lstlisting}

\subsection*{Self-disclosure}

\begin{lstlisting}
"You are a professional psychologist specializing in communication analysis and self-disclosure. Classify every utterance strictly according to the provided framework. If uncertain, make a best-effort classification."

"""
Task: Based on the session background, self-disclosure definition, and the client's utterance, classify the level of self-disclosure (G, M, or H) in the client's utterance and provide a brief explanation.

Self-Disclosure Definition:
Self-disclosure refers to the act of revealing personal, private, or sensitive information about oneself.
It can range from general statements with no personal relevance to deeply personal and emotional disclosures.

Three Self-disclosure levels:
1. G (General, No Disclosure)
    - The client talks about external topics without sharing personal experiences, thoughts, or emotions.
    - Example: "The weather has been nice lately."

2. M (Medium Disclosure)
    - The client shares non-sensitive personal experiences, plans, or general life updates. This includes age, occupation and hobbies, personal events, past history, future plans, or details about family members.
    - Example: "I started a new job last week."

3. H (High Disclosure)
    - The client reveals deeply personal, sensitive, or emotional experiences, including thoughts about mental health, relationships, or private struggles. This includes personal characteristics, problematic behaviors, physical appearance concerns, or wishful thoughts.
    - Example: "Rachel's been shutting everyone out lately. She barely leaves her room, and when she does, she hardly says a word. Tim keeps telling me to give her space, but... I can't shake this feeling that something's really wrong."

Session Background: {summary}
Client: "{utterance}"

Output Format:
Reasoning (<= 50 words): [justify your classification, citing specific elements of the utterance if applicable.]
Label:
"""
\end{lstlisting}

\subsection*{Emotion}

\begin{lstlisting}
"Act like an expert conversation evaluator with over 20 years of experience in assessing the emotions between clients and therapists. You are skilled in identifying subtle emotional cues and non-verbal indicators within dialogue. Your expertise includes analyzing speech patterns and emotional tone to evaluate the emotions. Your analysis must be thorough, evidence-based and objective."

"""
In this task, you will assess a therapy conversation between a client (C) and a therapist (T). Your goal is to analyze the central utterance using nine emotions, providing a rating for each emotion on a 5-point Likert scale.

You will analyze the conversation log for the following nine emotions:
1. Anger
2. Contempt
3. Disgust
4. Enjoyment
5. Fear
6. Sadness
7. Surprise
8. Anxiety
9. Depression

Use the following 5-point Likert scale to rate the intensity of each emotion:
1 - Very slightly or not at all,
2 - A little,
3 - Moderately,
4 - Quite a bit,
5 - Extremely.

Instructions:
1. Carefully analyze central utterance within the context of the preceding conversation.
2. For each of the nine emotions, identify specific conversational cues or statements that support your rating.
3. Provide a rating (1-5) for each emotion. If no clear emotion is present, assign a rating of 1 to all emotions.
4. Provide an evidence-based rationale for your rating, referencing contextual cues from the previous utterances. Avoid making assumptions not supported by the dialogue.

Contextual Analysis:
Use the utterances that precede the central utterance as the context (could be empty):
{context}

Analyze the following central utterance:
{text}

Take a deep breath and work on this problem step-by-step.
Your analysis:
"""
\end{lstlisting}

\subsection*{Rapport}

\begin{lstlisting}
"Act like an expert conversation evaluator with over 20 years of experience in assessing the therapeutic alliance between clients and therapists. You are skilled in identifying subtle conversational cues and non-verbal indicators within dialogue. Your expertise includes analyzing speech patterns, emotional tone, and implicit affirmations to evaluate the depth of the bond. Your analysis must be thorough, objective, and based on specific criteria using a detailed 7-point Likert scale."

"""
In this task, you will evaluate a conversation between a client (C) and a therapist (T). Your objective is to assess the quality of their bond using four key aspects, providing a rating for each aspect on a 7-point Likert scale. Additionally, give an overall rating for the entire session and offer brief, evidence-based explanations for each score.

You will analyze the conversation for the following four bond aspects:
1. There is a mutual liking between the client and therapist.
2. The client feels confident in the therapist's ability to help the client.
3. The client feels that the therapist appreciates him/her as a person.
4. There is mutual trust between the client and therapist.


Use the following 7-point Likert scale for each aspect:
1: "Very strong evidence against",
2: "Considerable evidence against",
3: "Some evidence against",
4: "No evidence",
5: "Some evidence",
6: "Considerable evidence",
7: "Very strong evidence".

Instructions:
1. Carefully read through the conversation log provided below.
2. For each of the four bond aspects, identify specific conversational cues or statements that support your rating.
3. Provide a rating (1-7) for each aspect and include a brief explanation, citing specific parts of the conversation that influenced your decision.
4. After rating each aspect, give an **overall bond rating** (1-7) for the entire conversation based on a holistic view of the interaction.
5. Your analysis should be objective, relying only on observable cues from the text. Avoid making assumptions not supported by the dialogue.

Now, based on four key aspects of bond, rate the provided conversation log on a 7-point Likert scale (1-7), considering mutual liking, confidence, appreciation, and trust, also considering subtle conversational cues, such as encouraging language and affirmations, that may indicate mutual respect or a sense of understanding:

Conversation log:
{conversation_log}

Provide an overall rating (1-7) for the entire conversation and with respect to each of the four key aspects of bond mentioned above. Briefly explain your rating.

Take a deep breath and work on this problem step-by-step.
Your rating:
"""
\end{lstlisting}

\section*{Appendix B: Additional Results}
\setcounter{table}{5}
\begin{table*}[tb]
\caption{This table summarizes the specific context window configurations used across constructs.}
\label{tab:context}
\small
\begin{tabularx}{\textwidth}{@{}P{0.18\textwidth}P{0.3\textwidth}L@{}}
\toprule
\textbf{Context Window Type} & \textbf{Constructs} & \textbf{Description} \\
\midrule
\textbf{2 prior client utterances} & Empathy - Emotional Reaction / Interpretation / Exploration / Reflection & Used for therapist-side evaluation; providing local grounding while minimizing noise. \\
\addlinespace
\textbf{2 prior utterances} & Self-Disclosure & Captures both the client’s evolving narrative and immediate response to therapist input, supporting accurate depth classification. \\
\addlinespace
\textbf{5 prior utterances} & Emotion & Empirically supported in prior work \cite{kojima2022large}. Broader context helps model subtle or evolving emotional states. \\
\addlinespace
\textbf{1 segment} & Rapport & Segment-level constructs inherently include multi-turn context and require no extra input. \\
\bottomrule
\end{tabularx}
\end{table*}
\begin{table*}[tb]
\caption{Prompt calibration datasets and model agreement for the three constructs with an external benchmark dataset; self-disclosure and emotion were calibrated against our own annotated validation sets.}
\label{tab:calibration}
\small
\begin{tabularx}{\textwidth}{@{}P{0.3\textwidth}P{0.27\textwidth}L@{}}
\toprule
\textbf{Construct} & \textbf{Evaluation Dataset} & \textbf{Model Performance} \\
\midrule
\textbf{Empathy} -- Emotional Reaction / Interpretation / Exploration & Empathy Mental Health Subreddit \cite{yang2024digital} & Pearson \(r = 0.77\) (pooled across 3 mechanisms); \(\bar{r} = 0.70\) \\
\addlinespace
\textbf{Empathy} -- Reflection & AnnoMI Dataset \cite{crawford2004panas} & F1 = 0.79 (reflection vs. non-reflection on 5 sessions) \\
\addlinespace
\textbf{Rapport} & 7 Cups WAI-O-S (Bond Subscale) \cite{kojima2022large} & Pearson \(r = 0.76\) (\(n = 51\) sessions; avg. of 4 human annotators) \\
\bottomrule
\end{tabularx}
\end{table*}
\begin{table*}[tb]
\caption{Intraclass correlation (ICC 2,k) between model-generated and human annotations, by construct family, for six large language models. Identical prompts were applied to every model with no model-specific tuning, and each was scored against the same human-annotated validation sets. Models are ordered by overall mean ICC.}
\label{tab:models}
\small
\begin{tabular}{@{}lllllll@{}}
\toprule
Model & Rapport & Emotion (mean) & Empathy (mean) & Reflection & Self-disclosure & Overall mean \\
\midrule
Gemini 3.5 Flash & 0.868 & 0.743 & 0.771 & 0.794 & 0.828 & 0.801 \\
GPT-5.5 & 0.842 & 0.717 & 0.800 & 0.789 & 0.809 & 0.791 \\
Claude Sonnet 4.6 & 0.832 & 0.682 & 0.844 & 0.779 & 0.813 & 0.790 \\
Gemini 3.1 Flash-Lite & 0.836 & 0.704 & 0.789 & 0.761 & 0.810 & 0.780 \\
Claude Opus 4.8 & 0.703 & 0.707 & 0.824 & 0.791 & 0.822 & 0.769 \\
GPT-4o-mini & 0.810 & 0.606 & 0.698 & 0.669 & 0.782 & 0.713 \\
\midrule
Cross-model range & 0.165 & 0.137 & 0.146 & 0.125 & 0.046 & 0.088 \\
Cross-model SD & 0.058 & 0.047 & 0.051 & 0.048 & 0.016 & 0.032 \\
\bottomrule
\end{tabular}

\vspace{4pt}
\begin{minipage}{\textwidth}\small
\emph{Note.} ICC (2,k), two-way random effects, absolute agreement, average measures. Emotion is the mean across the nine emotion categories; Empathy is the mean across emotional reaction, interpretation, and exploration. Overall mean is the unweighted mean of the five construct-family values. GPT-4o-mini was used for all measures in the original submission; in the revised analysis it is retained for all constructs except emotion, which was re-derived with Gemini 3.5 Flash. Model version strings and API access dates are reported in Table 9.
\end{minipage}
\end{table*}
\begingroup
\renewcommand{\thetable}{8b}
\begin{table*}[tb]
\caption{Agreement with human annotations for each emotion category under the original measurement model (GPT-4o-mini) and the replacement model (Gemini 3.5 Flash).}
\label{tab:emotion-models}
\begin{minipage}{\textwidth}\small
Emotions are ordered by the size of the improvement. The four largest gains are the four categories that constitute the self-directed negative emotion composite.
\end{minipage}

\vspace{6pt}
\small
\begin{tabular}{@{}llll@{}}
\toprule
\textbf{Emotion} & \textbf{GPT-4o-mini ICC (95\% CI)} & \textbf{Gemini 3.5 Flash ICC (95\% CI)} & \textbf{Δ ICC} \\
\midrule
Anxiety & 0.465 (−0.19 to 0.74) & 0.748 (0.695 to 0.792) & \textbf{+0.283} \\
Fear & 0.453 (−0.13 to 0.71) & 0.706 (0.626 to 0.768) & \textbf{+0.253} \\
Sadness & 0.524 (−0.16 to 0.77) & 0.733 (0.676 to 0.782) & \textbf{+0.209} \\
Depression & 0.489 (−0.06 to 0.72) & 0.659 (0.591 to 0.719) & \textbf{+0.170} \\
Enjoyment & 0.686 (0.43 to 0.81) & 0.798 (0.732 to 0.849) & +0.112 \\
Disgust & 0.723 (0.58 to 0.81) & 0.799 (0.744 to 0.842) & +0.076 \\
Surprise & 0.621 (0.52 to 0.70) & 0.677 (0.583 to 0.749) & +0.056 \\
Anger & 0.745 (0.57 to 0.83) & 0.796 (0.745 to 0.837) & +0.051 \\
Contempt & 0.745 (0.66 to 0.80) & 0.771 (0.718 to 0.812) & +0.026 \\
\midrule
\textbf{Mean} & \textbf{0.606} & \textbf{0.743} & \textbf{+0.137} \\
\bottomrule
\end{tabular}
\end{table*}
\endgroup
\addtocounter{table}{-1}
\begin{table*}[tb]
\caption{Vendor, Model, and API Access Date}
\label{tab:vendors}
\small
\centering
\begin{tabular}{@{}lll@{}}
\toprule
\textbf{Vendor} & \textbf{Model} & \textbf{API Access Date} \\
\midrule
OpenAI & GPT-4o-mini & January 2025 \\
OpenAI & GPT-5.5 & June 2026 \\
Anthropic & Claude Sonnet 4.6 & June 2026 \\
Anthropic & Claude Opus 4.8 & June 2026 \\
Google & Gemini 3.1 Flash-Lite & June 2026 \\
Google & Gemini 3.5 Flash & June 2026 \\
\bottomrule
\end{tabular}
\end{table*}
\begin{table*}[tb]
\caption{Correlations among the five client-level outcomes}
\label{tab:outcome-corr}
\small
\centering
\begin{tabular}{@{}llllll@{}}
\toprule
\textbf{Outcome} & \textbf{1} & \textbf{2} & \textbf{3} & \textbf{4} & \textbf{5} \\
\midrule
1. Disclosure & --- & & & & \\
2. Self-directed negative emotion & .577 & --- & & & \\
3. Outward-directed negative emotion & .238 & .229 & --- & & \\
4. Surprise & .062 & .033 & .095 & --- & \\
5. Enjoyment & −.035 & −.253 & −.100 & .102 & --- \\
\bottomrule
\end{tabular}
\end{table*}
\begin{table*}[tb]
\caption{Correlations among the four predictors}
\label{tab:predictor-corr}
\small
\centering
\begin{tabular}{@{}lllll@{}}
\toprule
\textbf{Predictor} & \textbf{1} & \textbf{2} & \textbf{3} & \textbf{4} \\
\midrule
1. Empathy & --- & & & \\
2. Exploration & .134 & --- & & \\
3. Rapport (prior session) & −.002 & .006 & --- & \\
4. log Session ID & .008 & −.002 & −.076 & --- \\
\bottomrule
\end{tabular}
\end{table*}
\begin{table*}[tb]
\caption{Estimates from multilevel models with random intercepts for session and client. Standardized coefficients with 95\% confidence intervals.}
\label{tab:mlm}
\setlength{\tabcolsep}{4pt}
\small
\begin{tabular}{@{}llllll@{}}
\toprule
\textbf{Predictor} & \textbf{\makecell[tl]{Self-directed \\ negative emotion}} & \textbf{\makecell[tl]{Outward-directed \\ negative emotion}} & \textbf{Surprise} & \textbf{Enjoyment} & \textbf{Disclosure} \\
\midrule
log Session ID & \makecell[tl]{0.007 \\ \relax (−0.008 to 0.022)} & \makecell[tl]{0.015 \\ \relax (−0.001 to 0.030)} & \makecell[tl]{0.006 \\ \relax (−0.003 to 0.015)} & \makecell[tl]{−0.002 \\ \relax (−0.015 to 0.012)} & \makecell[tl]{0.012 \\ \relax (−0.001 to 0.024)} \\
\addlinespace
Rapport (prior session) & \makecell[tl]{−0.020 \\ \relax (−0.037 to −0.002)} & \makecell[tl]{−0.009 \\ \relax (−0.027 to 0.009)} & \makecell[tl]{−0.001 \\ \relax (−0.011 to 0.009)} & \makecell[tl]{0.016 \\ \relax (0.001 to 0.032)} & \makecell[tl]{−0.003 \\ \relax (−0.017 to 0.012)} \\
\addlinespace
Exploration & \makecell[tl]{0.065 \\ \relax (0.059 to 0.070)} & \makecell[tl]{−0.007 \\ \relax (−0.013 to −0.002)} & \makecell[tl]{−0.015 \\ \relax (−0.021 to −0.009)} & \makecell[tl]{−0.040 \\ \relax (−0.046 to −0.034)} & \makecell[tl]{0.072 \\ \relax (0.066 to 0.077)} \\
\addlinespace
Empathy & \makecell[tl]{0.097 \\ \relax (0.092 to 0.103)} & \makecell[tl]{0.009 \\ \relax (0.004 to 0.015)} & \makecell[tl]{0.006 \\ \relax (0.000 to 0.012)} & \makecell[tl]{−0.036 \\ \relax (−0.041 to −0.030)} & \makecell[tl]{0.064 \\ \relax (0.058 to 0.070)} \\
\bottomrule
\end{tabular}

\vspace{8pt}
\textbf{Exact \emph{P} values}

\vspace{4pt}
\begin{tabular}{@{}llllll@{}}
\toprule
\textbf{Predictor} & \textbf{Self-directed} & \textbf{Outward-directed} & \textbf{Surprise} & \textbf{Enjoyment} & \textbf{Disclosure} \\
\midrule
log Session ID & .37 & .06 & .17 & .83 & .07 \\
Rapport & .03 & .32 & .82 & .04 & .72 \\
Exploration & <.001 & .01 & <.001 & <.001 & <.001 \\
Empathy & <.001 & .001 & .045 & <.001 & <.001 \\
\bottomrule
\end{tabular}

\vspace{4pt}
\begin{minipage}{\textwidth}\small
\emph{Note.} N = 119,233 client utterances from 1,573 sessions and 37 clients. Estimated by restricted maximum likelihood with random intercepts for session and for client. All predictors and outcomes are standardized within client, so coefficients are within-client standardized associations. Emotion outcomes were derived with Gemini 3.5 Flash; therapist behavior and disclosure measures with GPT-4o-mini. Neither rapport association is robust across specifications.
\end{minipage}
\end{table*}
\begin{table*}[tb]
\caption{Variance decomposition and explained variance by outcome.}
\label{tab:variance}
\setlength{\tabcolsep}{4pt}
\small
\begin{tabular}{@{}lllll@{}}
\toprule
\textbf{Outcome} & \textbf{Session-level variance} & \textbf{Residual (utterance-level)} & \textbf{Marginal R²} & \textbf{Conditional R²} \\
\midrule
Self-directed negative emotion & 10.5\% & 89.5\% & 0.016 & 0.119 \\
Outward-directed negative emotion & 10.5\% & 89.5\% & 0.001 & 0.106 \\
Surprise & 2.2\% & 97.8\% & <0.001 & 0.023 \\
Enjoyment & 7.5\% & 92.5\% & 0.004 & 0.078 \\
Disclosure & 6.4\% & 93.6\% & 0.011 & 0.074 \\
\bottomrule
\end{tabular}

\vspace{4pt}
\begin{minipage}{\textwidth}\small
\emph{Note.} Client-level variance was below 0.05\% for all outcomes and is not shown separately. Marginal R² reflects the fixed effects alone; conditional R² includes the random effects (Nakagawa method). The gap between the two columns is the variance attributable to stable between-session differences rather than to the therapist behaviors modeled here.
\end{minipage}
\end{table*}
\begin{table*}[tb]
\caption{Fit statistics for the multilevel models, and comparison against single-level models with identical fixed effects.}
\label{tab:fit}
\setlength{\tabcolsep}{4pt}
\small
\begin{tabular}{@{}lllllll@{}}
\toprule
\textbf{Outcome} & \textbf{AIC (single-level)} & \textbf{AIC (two-level)} & \textbf{ΔAIC} & \textbf{LR (random intercept)} & \textbf{ICC} & \textbf{LR (predictors, 4 df)} \\
\midrule
Self-directed negative emotion & 335,593 & 327,088 & −8,505 & 8,506.8 & 0.105 & 1,909.4 \\
Outward-directed negative emotion & 339,232 & 332,050 & −7,182 & 7,184.6 & 0.105 & 20.0 \\
Surprise & 338,803 & 337,728 & −1,075 & 1,076.9 & 0.022 & 29.6 \\
Enjoyment & 338,440 & 332,895 & −5,544 & 5,546.3 & 0.075 & 392.3 \\
Disclosure & 336,201 & 331,791 & −4,410 & 4,412.3 & 0.064 & 1,245.5 \\
\bottomrule
\end{tabular}

\vspace{4pt}
\begin{minipage}{\textwidth}\small
\emph{Note.} All likelihood ratio tests \emph{P} < .001. The random-intercept test compares the two-level model against a single-level model with identical fixed effects; because the null places the variance parameter on the boundary of the parameter space, \emph{P} values are computed against a 50:50 mixture of \(\chi^2_0\) and \(\chi^2_1\) rather than a naive \(\chi^2_1\). The predictor test compares the full model against an intercept-only multilevel model (maximum likelihood fits). With N = 119,233, these tests are highly sensitive --- the outward-directed model reaches \emph{P} < .001 with a marginal R² of 0.001 --- so the R² values in Table 13 are the more informative summary of explanatory power.
\end{minipage}
\end{table*}

\end{document}